\documentclass[pre,superscriptaddress,showpacs]{revtex4}

\usepackage{graphicx}%
\usepackage{dcolumn}
\usepackage{amsmath}

\usepackage[dvips]{epsfig}
\usepackage{amssymb}
\usepackage{array,longtable}
\usepackage{amscd}

\begin{document}



\newcommand{\be}{\begin{eqnarray}}
\newcommand{\ee}{\end{eqnarray}}
\newcommand{\bse}{\begin{subequations}}
\newcommand{\ese}{\end{subequations}}

\newcommand{\bs}{\boldsymbol}
\newcommand{\mbb}{\mathbb}
\newcommand{\mcal}{\mathcal}
\newcommand{\mfr}{\mathfrak}
\newcommand{\mrm}{\mathrm}

\newcommand{\ovl}{\overline}
\newcommand{\p}{\partial}
\newcommand{\f}{\frac}
\newcommand{\diff}{\mrm{d}}
\newcommand{\lan}{\langle}
\newcommand{\ran}{\rangle}

\newcommand{\ga}{\alpha}
\newcommand{\gb}{\beta}
\newcommand{\gc}{\gamma}
\newcommand{\gd}{\delta}
\newcommand{\Gc}{\Gamma}
\newcommand{\gl}{\lambda}
\newcommand{\gk}{\kappa}
\newcommand{\go}{\omega}
\newcommand{\Go}{\Omega}
\newcommand{\veps}{\varepsilon}

\newcommand{\sn}{\mrm{sn}}
\newcommand{\cn}{\mrm{cn}}
\newcommand{\dn}{\mrm{dn}}
\newcommand{\am}{\mrm{am}}
\newcommand{\sech}{\mrm{sech}}
\newcommand{\sign}{\mrm{sign}}

\newcommand{\csp}{\;,\qquad\qquad}
\newcommand{\fa}{\forall\;}

\newcommand{\N}{\mbb{N}}
\newcommand{\R}{\mbb{R}}
\newcommand{\D}{\mcal{D}}
\newcommand{\Nn}{\mcal{N}}

\newcommand{\im}{\mrm{image}\;}
\newcommand{\num}{\mrm{\#}}


\title{Theory of the Relativistic Brownian Motion. The 1+1-Dimensional Case}

\author{J\"orn Dunkel}
\email{dunkel@physik.hu-berlin.de}
\affiliation{Institut f\"ur Physik, Humboldt-Universit\"at zu Berlin,
 Newtonstra{\ss}e 15, D-12489 Berlin, Germany}

\author{Peter H\"anggi}
\affiliation{Institut f\"ur Physik, Universit\"at Augsburg,
 Theoretische Physik I,  Universit\"atstra{\ss}e 1, D-86135 Augsburg, Germany}

\date{\today}

\begin{abstract}
We construct a theory for the 1+1-dimensional Brownian motion in a
viscous medium, which is (i) consistent with Einstein's theory of
special relativity, and (ii) reduces to the standard Brownian motion
in the Newtonian limit case. In the first part of this work the
classical Langevin equations of motion, governing the
nonrelativistic dynamics of a free Brownian particle in the presence
of a heat bath (white noise), are generalized in the framework of
special relativity. Subsequently, the corresponding relativistic
Langevin equations are discussed in the context of the generalized Ito
(pre-point discretization rule) {\it vs.} the Stratonovich
(mid-point discretization rule)  dilemma: It is found that the
relativistic Langevin equation in the H\"anggi-Klimontovich
interpretation (with the post-point discretization rule) is the only
one that yields agreement with the relativistic Maxwell
distribution. Numerical results for the relativistic Langevin
equation of a free Brownian particle are presented.
\end{abstract}


\maketitle

\section{Introduction}
\label{introduction}

For almost one-hundred years, Einstein's theory of special
relativity \cite{Ei05a,Ei05b} is serving as the foundation of our
most successful physical standard models (apart from gravity). The
most prominent and, probably, also the most important feature of
this theory is the absolute character of the speed of light $c$,
representing an unsurmountable barrier for the velocity of any
(macroscopic) physical process. Due to the great experimental
success of the original theory,  almost all other physical theories
have successfully been adapted to the framework of special
relativity over the past decades. Surprisingly, however, the
scientific literature provides relatively few publications on the
subject of relativistic Brownian motions (classical references are
\cite{Sc61,Ha65,Du65}, more recent contributions include \cite{GuRu78,Bo79a,Bo79b,BY81,MoVi95,Po97,OrHo03,FrLJ04}).
\par
Brownian particles are physical objects (e.g., dust grains), which
move randomly through a surrounding medium (heat bath). Their
stochastic motions are caused by permanent collisions with  much
lighter constituents  of the heat bath (e.g., molecules of a
liquid). The classical theory of Brownian motion or {\it
nonrelativistic} diffusion theory, respectively, was developed by
Albert Einstein \cite{Ei05c} and Marian von Smoluchowski
\cite{EiSm}. Since the beginning of the last century, when their
seminal papers were published, the classical theory has been
investigated and generalized by a large number of physicists
\cite{WaUh45,UhOr30,Ch43,HaTo82,VK03} and mathematicians
\cite{Wi23,Gr02,KaSh91}. The intense research led, among others, to
different mathematical representations of the Brownian motion
dynamics (Langevin equations, Fokker-Planck equations, etc.
\cite{Ch43,HaTo82,VK03}), to the notion of Wiener processes
\cite{Wi23}, and to new techniques for solving partial differential
equations (Feynman-Kac formula, etc. \cite{Gr02,KaSh91}).
\par
With regard to special relativity, standard Brownian motion faces the problem
that it permits velocity jumps $\Delta v$, that exceed the speed of light $c$
(see also Schay \cite{Sc61}). This is due to the fact that in the
nonrelativistic theory the velocity increments  $\Delta v$ have a Gaussian
distribution, which always assigns a nonvanishing (though small) probability to
  events $\Delta v >c$. This problem is also reflected by the Maxwell
distribution, which represents the stationary velocity distribution for
an ensemble of free Brownian particles and permits absolute
velocity values $v>c$ \cite{VK03}.
\par
The first relativistically consistent generalization of Maxwell's
velocity distribution was introduced by J\"uttner \cite{Ju11} in
1911. Starting from an extremum principle for the entropy, he
obtained the probability distribution function of the relativistic
ideal Boltzmann gas (see Eq. \eqref{e:juettner} below). In
principle, however, J\"uttner's approach made no contact with the
theory of Brownian motion. Fifty years after J\"uttner's work, Schay
\cite{Sc61} performed the first comprehensive mathematical
investigation of relativistic diffusion processes, based on
Lorentz-invariant transition probabilities. On the mathematical
side, Schay's analysis was complemented by Hakim \cite{Ha65} and Dudley
\cite{Du65}, who studied in detail the properties of Lorentz-invariant Markov
processes in relativistic phase space. After forty more years,
Franchi and Le Jan \cite{FrLJ04} have recently presented an
extension of Dudley's work to general relativity. In particular,
these authors discuss relativistic diffusions in the presence of a
Schwarzschild metric \cite{Weinberg}. Hence, over the past
one-hundred years there has been steady (though relatively slow)
progress in the mathematical analysis of relativistic diffusion
processes.
\par
By contrast, one finds in the physical literature only very few
publications which directly address the topic of the relativistic
Brownian motion (despite the fact that relativistic kinetic theory
is fairly well established for more than thirty years
\cite{Li66,St71,DG80,Liboff90}). Among the few exceptions are the papers by
Boyer \cite{Bo79a,Bo79b} and Ben-Ya'acov \cite{BY81}, who have studied the
interaction between two energy-level particles and electromagnetic
radiation in thermal equilibrium, the latter acting as a heat bath.
In contrast to their specific microscopic model, we shall adopt a
more coarse-grained point of view here by assuming that the heat
bath is sufficiently well described by macroscopic friction and
diffusion coefficients.
\par
Generally, the objective of the present paper can be summarized as
follows: We would like to discuss how one can construct, in a
physically straightforward manner, a relativistic theory of Brownian
motion for particles moving in a homogeneous, viscous medium. For
this purpose it is sufficient to concentrate on the case of
1+1-dimensions (generalizations to the 1+3-dimensions are
straightforward and will be discussed separately in a forthcoming
contribution). As a starting point we choose the nonrelativistic
Langevin equations of the free Brownian particle. In Sec.
\ref{langevin_approach} these equations will be generalized such
that they comply with special relativity. As we shall see in Sec.
\ref{FPE}, due to multiplicative noise for the momentum degree of
freedom, the resulting relativistic Langevin equations are {\em not}
sufficient in order to uniquely determine the corresponding
Fokker-Planck equation (generalized Ito-Stratonovich dilemma).
Furthermore, it it is shown that the stationary solution of a
particular form for the relativistic Fokker-Planck equation
coincides with J\"uttner's relativistic Maxwell distribution (Sec.
\ref{FPE-jutt}). Finally, we also discuss numerical results for the
mean square displacement in Sec. \ref{numerics}.
\par
It might be worthwhile to emphasize that the systematic Langevin approach
 pursued below is methodically different from those in
 Refs. \cite{GuRu78,Bo79a,Bo79b,BY81,MoVi95,Po97,OrHo03,Sc61,Ha65,Du65,FrLJ04}
 and also from the kinetic theory approach \cite{Li66,St71,DG80,Liboff90}. It
 is therefore satisfactory that our findings are apparently consistent with
 rigorous mathematical results, obtained by Schay \cite{Sc61} and Dudley
 \cite{Du65} for the case of free relativistic diffusion. Moreover, it will
 become clear in Sec. \ref{numerics} that numerical simulations of the
 relativistic Langevin equations constitute a very useful tool for the
 numerical investigation of relativistic diffusion processes, provided that
 the discretization rule is carefully chosen.

\section{Langevin dynamics}
\label{langevin_approach}

First the main properties of the nonrelativistic Langevin equations for free Brownian particles are briefly summarized
(Sec. \ref{physical_foundations}). Subsequently, we construct generalized Lorentz-covariant Langevin equations
(Sec. \ref{relativistic_generalization}). Finally, the covariant Langevin equations will be rewritten in laboratory coordinates
(Sec. \ref{lab}).
\par
The following notations will be used throughout the paper: Since we confine
ourselves to the 1+1-dimensional case, upper and lower
Greek indices $\ga,\gb,\ldots$ can take values $0,1$, where \lq$0$\rq\space
refers to the time component. The  1+1-dimensional Minkowski metric tensor with respect to Cartesian coordinates is taken as
$$
(\eta_{\ga\gb})=(\eta^{\ga\gb})=\mrm{diag}(-1,1).
$$
Moreover, Einstein's summation convention is invoked throughout.

\subsection{Physical foundations}
\label{physical_foundations}
Consider the nonrelativistic one-dimensional motion of a Brownian particle
with mass $m$ that is surrounded by a heat bath (e.g., small liquid
particles). In the Langevin approach the nonrelativistic dynamics of the
Brownian particle is described by the stochastic dynamical equations (see, e.g., \cite{VK03} Chap. IX)
\bse\label{e:langevin}
\be
\f{\diff x}{\diff t}(t) &=&\label{e:langevin_a}
v(t)\\
m\f{\diff v}{\diff t}(t)&=& \label{e:langevin_b} -\nu\, m  v(t)+ L(t),
\ee
\ese
where $\nu$ is the viscous friction coefficient. The Langevin
force $L(t)$ is characterized by \be\label{e:white_noise} \lan
L(t)\ran = 0,\qquad\qquad \lan L(t) L(s) \ran= 2 D\, \gd(t-s), \ee
with all higher cumulants being zero (Gaussian white noise), and $D$
being constant. More general models may include velocity-dependent
parameters $\nu$ and $D$ (see, e.g., \cite{ErEbScSc00,HaTo82,Kl94,Tr03}), but we shall
restrict ourselves to the simplest case here. It is worthwhile to
summarize the physical assumptions, implicitly underlying Eqs.
\eqref{e:langevin}:
\begin{itemize}
\item the heat bath is homogeneous,
\item stochastic impacts between the Brownian particle and the constituents of the heat bath occur virtually uncorrelated,
\item on the macroscopic level, the interaction between Brownian particle and heat
bath is sufficiently well described by the constant viscous friction coefficient $\nu$ and the white noise force $L$,
\item Eqs. \eqref{e:langevin} hold in the rest frame $\Sigma_0$ of the heat bath
  (corresponding to the specific inertial system, in which the average
  velocity of the heat bath vanishes for all times $t$).
\end{itemize}
In the following $\Sigma_0$ will also be referred to as {\em laboratory frame}.
\par
In the mathematical literature, Eq. \eqref{e:langevin_b} is usually
written as \bse\label{e:langevin_m} \be\label{e:langevin_math}
\diff\left[m v(t)\right]=-\nu\, m v(t)\diff t+\diff W(t), \ee where $W(t)$ is
a one-dimensional Wiener process \cite{HaTo82,KaSh91,Gr02}, i.e.,
the density of the increments \be w(t)\equiv \diff W(t)\equiv W(t+\diff
t)-W(t) \ee is given by \be\label{e:langevin_math_density}
\mcal{P}^1[w(t)]=\f{1}{\sqrt{4\pi D\; \diff t}}
\exp\left[-\f{w(t)^2}{4D\,\diff t}\right]. \ee \ese
Here the
abbreviation $w\equiv \diff W$ has been introduced to simplify the
notation in subsequent formulae. From Eq.
\eqref{e:langevin_math_density} one finds in agreement with
\eqref{e:white_noise} \be\label{e:white_noise_math} \lan w(t)\ran =
0,\qquad\qquad \lan w(t)\, w(s) \ran=
\begin{cases}
0, & t\ne s\\
2 D\,\diff t, &t=s.
\end{cases}
\ee
Depending on which notation is more convenient for the current purpose, we
shall use below either the physical formulation \eqref{e:langevin} or the
mathematical formulation \eqref{e:langevin_m}. The two formulations can
 be connected by (formally) setting
\be
w(t)=\diff W(t)= L(t)\diff t.
\ee

\subsection{Relativistic generalization}
\label{relativistic_generalization}
It is well-known that in inertial coordinate systems, which are
comoving with a particle at a given moment $t$, the relativistic equations
must reduce to the nonrelativistic Newtonian equations (see, e.g.,
\cite{Weinberg} Chap. 2.3). Therefore our strategy is as follows:
Starting from the Langevin equations \eqref{e:langevin} or \eqref{e:langevin_math}, respectively, we
construct in the first step the nonrelativistic equations of motion with
respect to a  coordinate frame $\Sigma_*$, comoving with the Brownian
particle at a given moment $t$. In the second step, the general form of
the covariant relativistic equations motions are found by applying a Lorentz
transformation to the nonrelativistic equations that have been obtained for
$\Sigma_*$.
\par
It is useful to begin by considering the deterministic (noise-free) limit case,
corresponding to a pure damping of the particle's motion. This will be done
Sec. \ref{viscous_friction}. Subsequently, the stochastic force is
separately treated in Sec. \ref{stochastic_force}.

\subsubsection{Viscous friction}
\label{viscous_friction}

Setting the stochastic force term to zero (corresponding to a
vanishing temperature of the heat bath), the nonrelativistic Eq.
\eqref{e:langevin_b} simplifies to
\be\label{e:langevin_friction}
m\f{\diff v}{\diff t} (t)=-\nu\, m  v(t).
\ee
The energy of the Brownian particle is purely kinetic,
\be
E(t)=\f{m v(t)^2 }{2},
\ee
and, by virtue of \eqref{e:langevin_friction}, its time derivative is given
by
\be\label{e:langevin_friction_energy}
\f{\diff E}{\diff t}  =m
v\f{\diff v}{\diff t} =-\nu\, m v^2.
\ee
As stated above, in the nonrelativistic theory the last three equations are assumed to hold
in the rest frame  $\Sigma_0$ of the heat bath. Now consider another
inertial coordinate system $\Sigma_*$, in which the Brownian
particle is temporarily at rest at time $t$ or $t_*=t_*(t)$, respectively,
where $t_*$ denotes the $\Sigma_*$-time coordinate. That is, in  $\Sigma_*$ we
have at time $t$
\be\label{e:comoving}
v_*(t)\equiv v_*\bigl(t_*(t)\bigr)=0.
\ee
(Conventionally, we use throughout the lax notation $g_*(t)\equiv
g_*(t_*(t))$, where $g_*$ is originally a function of $t_*$.)
With respect to the comoving frame $\Sigma_*$, the heat bath will,
in general, have a non-vanishing (average) velocity $V_*$. Then,
using a Galilean transformation we find that Eq.
\eqref{e:langevin_friction}  in $\Sigma_*$-coordinates at time $t$
reads as follows
\bse\label{e:EOM_comoving}
\be
m\f{\diff v_*}{\diff t_*}(t)=-\nu\, m [v_*(t)- V_*]
\overset{\eqref{e:comoving}}{=}\nu\, m  V_*.
\ee
Similarly, in $\Sigma_*$-coordinates Eq.
\eqref{e:langevin_friction_energy} is given by
\be \f{\diff E_*}{\diff t_*}
(t)=-\nu\, m v_*(t)\left[v_*(t)- V_*\right]
\overset{\eqref{e:comoving}}{=}0.
\ee
\ese
Note that in the nonrelativistic (Newtonian) theory the left equalities in Eqs.
\eqref{e:EOM_comoving} are valid for arbitrary time $t$. By
contrast, in the relativistic theory these equations are exact at
time $t$ only if $\Sigma_*$ is comoving at time $t$. In the latter
case, we can use Eqs. \eqref{e:EOM_comoving} to construct
relativistically covariant equations of motion. Introducing, as
usual, the proper time $\tau$ by the definition
\be \label{e:propertime}
\diff \tau\equiv
\diff t \sqrt{1-\f{v^2}{c^2}}=\diff t_* \sqrt{1-\f{v_*^2}{c^2}},
\ee
and combining momentum $p_*=m v_*$ and
energy into a 1+1-vector $(p_*^\ga)=(p^0,p_*)=(E_*/c,p_*)$, we can
rewrite Eqs. \eqref{e:EOM_comoving} in the covariant form
\be\label{e:EOM_friction_covariant.1}
\f{\diff p^\ga_*}{\diff \tau}
=f^\ga_*, \qquad\qquad (f^\ga_*)=-m \nu\left( 0,\; v_*-
V_*\right).
\ee
Let $(u^\ga_*)$ and $(U^\ga_*)$ denote the 1+1-velocity components of Brownian particle and heat bath,
respectively. Now it is important to realize that the covariant
force vector $f^\ga$ can {\em not} be simply proportional to the
1+1-velocity difference,
\be
f_*^\ga\ne -m\nu (u_*^\ga-U_*^\ga),
\ee
since, in general, at time $t$ in $\Sigma_*$
\be
u_*^0-U_*^0
=\f{c}{\sqrt{1-v_*^2/c^2}}-\f{c}{\sqrt{1-V_*^2/c^2}}
\overset{\eqref{e:comoving}}{=}{c}-\f{c}{\sqrt{1-V_*^2/c^2}} \ne 0.
\ee
However, we can write $f_*^\ga$ in a manifestly covariant form,
if we introduce the {\em friction tensor}
\be\label{e:friction_tensor} ({{\nu_*}^\ga}_\gb)=\begin{pmatrix}
0&0\\
0&\nu
\end{pmatrix},
\ee
which allows us to rewrite \eqref{e:EOM_friction_covariant.1} as
\be\label{e:EOM_friction_covariant}
\f{\diff p^\ga_*}{\diff \tau}= -m\,{{\nu_*}^\ga}_\gb (u_*^\gb-U_*^\gb).
\ee
This equation is manifestly Lorentz-invariant and we drop the asterisk from
now on, while keeping in mind that the diagonal form of the friction tensor
\eqref{e:friction_tensor} is linked to the rest frame $\Sigma_*$ of the
Brownian particle. In this respect the friction tensor is very similar to the
pressure tensor, as known from the relativistic hydrodynamics of perfect
fluids (see, e.g., \cite{Weinberg} Chap. 2.10). This analogy yields
immediately the following
representation
\be\label{e:friction_tensor_covariant}
{{\nu}^\ga}_\gb= \nu\left({\eta^\ga}_{\gb}+\f{u^\ga u_\gb}{c^2}\right).
\ee
\par
It is now interesting to consider Eq. \eqref{e:EOM_friction_covariant} in the laboratory
frame $\Sigma_0$, defined above as the rest frame of
the heat bath.  There we have
\be\label{e:rest_frame_heat}
(U^\gb)=(c, 0),\qquad\qquad
(u^\gb)= (\gc c,\gc v),\qquad\qquad
\diff \tau =\f{\diff t}{\gc},\qquad\qquad
\gc \equiv \f{1}{\sqrt{1- v^2/c^2}}
\ee
Combining \eqref{e:EOM_friction_covariant},
\eqref{e:friction_tensor_covariant}, and \eqref{e:rest_frame_heat}
we find that the relativistic equations of motion in
$\Sigma_0$ are given by
\bse\label{e:EOM_lab}
\be
\f{\diff p}{\diff t}&=&\label{e:EOM_lab_1}
-\nu \f{m  v}{\sqrt{1- v^2/c^2}}\\
\f{\diff E}{\diff t}&=&\label{e:EOM_lab_2}
-\nu  \f{m v^2}{\sqrt{1- v^2/c^2}}.
\ee
\ese
Upon comparing \eqref{e:EOM_lab_1} with \eqref{e:langevin_friction} and
\eqref{e:EOM_lab_2}  with \eqref{e:langevin_friction_energy}, one readily
observes that the
relativistic equations \eqref{e:EOM_lab} do indeed reduce to the known
Newtonian laws in the limit case $v^2/c^2\ll 1$.
\par
\begin{figure}[h]
\center
\epsfig{file=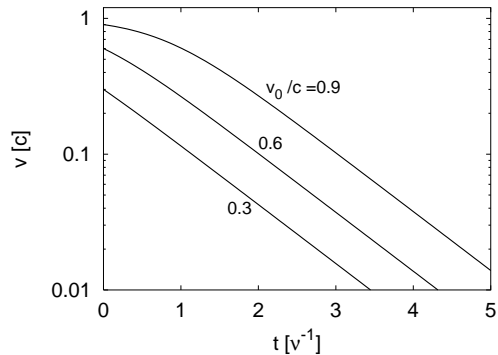,height=7cm, angle=-90}
\caption{Velocity curves $v(t)$, corresponding to Eq. \eqref{e:velocity_pure_friction},
for the purely damped motion of a relativistic
  particle in the rest frame of the heat bath (laboratory frame). Especially
  at high velocities $|v| \lesssim c$, the relativistic velocity curves deviate from
  the exponential decay, predicted by the Newtonian theory. \label{fig01}}
\end{figure}

Using the relativistic definitions
\be\label{e:definition_relativistic_momentum}
E=\gc mc^2,\qquad\qquad
{p}=\gc m v,
\ee
Eqs. \eqref{e:EOM_lab} can also be rewritten as
\bse\label{e:EOM_lab_rew}
\be
\f{\diff p}{\diff t}&=&\label{e:EOM_lab_rew-1}
-\nu\, p,\\
\f{\diff E}{\diff t}&=&-\nu\,  p\, v=-\nu E \f{ v^2}{c^2}
\ee
\ese
In fact, only one of the two Eqs. \eqref{e:EOM_lab} or
\eqref{e:EOM_lab_rew}, respectively, must be solved, due to the fixed relation
between relativistic energy and momentum:
\be
p_\ga p^\ga=-E^2/c^2+ p^2=-m^2c^2
\qquad \Rightarrow\qquad
E(t)=\f{mc^2}{\sqrt{1- v^2/c^2}}.
\ee
The solution of \eqref{e:EOM_lab_rew-1} reads
\be
{p}(t)={p}_0 \exp(-\nu t),
\qquad\qquad
{p}(0)={p}_0,
\ee
and, by using \eqref{e:definition_relativistic_momentum}, one thus obtains for the
velocity of the particle in the laboratory frame $\Sigma_0$ (rest frame
of the heat bath)
\be\label{e:velocity_pure_friction}
 v(t)= v_0
\left[\left(1-\f{v_0^2}{c^2}\right)e^{2\nu
t}+\f{v_0^2}{c^2}\right]^{-1/2}.
\ee
Figure \ref{fig01} depicts a
semi-logarithmic representation of the velocity $v(t)$ for different
values of the initial velocity $v_0$. As one can see in the diagram,
at high velocities $|v| \lesssim c$  the relativistic velocity
curves, given by Eq. \eqref{e:velocity_pure_friction}, exhibit
essential deviations from the purely exponential decay, predicted by
the Newtonian theory.

\subsubsection{Stochastic force}
\label{stochastic_force}

We now construct a {\it relativistic generalization of the stochastic force}.  To
this end, we consider Eq. \eqref{e:propertime} as an {\em operational
definition} for the proper time parameter $\tau$. The generalization procedure will be based on the standard assumption (postulate)
that, in temporarily comoving inertial frames $\Sigma_*$, the relativistic equations of
motions must reduce to the Newtonian equations of motions. According to this
assumption, for frames $\Sigma_*$, comoving with the particle at laboratory time $t$, the relativistic stochastic
differential equation must reduce to
\be\label{e:langevin_math_comoving}
\diff p_*(t)=-\nu\left[p_*(t)-m V_*\right]\diff t_*+ w_*(t),
\ee
where the momentum increments $w_*\equiv\diff W_*$ represent a Wiener-process with parameter $D$, i.e., the increments $w_*(t)$ have a Gaussian distribution:
\be\label{e:wiener}
\mcal{P}_*^1[ w_*(t)]=
\f{1}{\sqrt{4\pi D\; \diff t_*}}
\exp\left[-\f{w_*(t)^2}{4D\,\diff t_*}\right].
\ee
Note that also in the relativistic theory the {\em momentum} increments
$w(t)=\diff W(t)$ may tend to infinity, as long as the related  {\em velocity}
increments remain bounded. In other words, in the relativistic theory one must
carefully distinguish between stochastic momentum and velocity increments
(this is not necessary in the nonrelativistic theory, since Newtonian momenta
are simply proportional to their velocities).
\par
The next step is now to define the {\em increment 1+1-vector} by
\be\label{e:covariant_stochastic_force}
(w^\ga_*)=(0,w_*).
\ee
This definition is in agreement with the requirement that in a comoving inertial system
$\Sigma_*$ the $0$-component of the 1+1-force vector must vanish
(see, e.g., \cite{Weinberg} Chap. 2.3, and also compare Eqs. \eqref{e:EOM_friction_covariant.1},
\eqref{e:condition.1} and \eqref{e:condition.2} of the present paper). Moreover, if the
Lorentz frame $\Sigma_*$ is comoving with the Brownian particle at given time
$t$, then the (equal-time) white noise relations \eqref{e:white_noise_math} generalize to
\be\label{e:relativistic_white_noise_math}
\lan w_*^\ga(t)\ran = 0,\qquad\qquad
\lan w_*^\ga(t)\, w_*^\gb(t) \ran=
\begin{cases}
0,                  & \ga=0\quad\mrm{and/or}\quad \gb=0,\\
2 D\,\diff t_*\;, &  \mrm{otherwise}.
\end{cases}
\ee
The rhs. of the second equation in \eqref{e:relativistic_white_noise_math} makes it plausible to introduce a {\em correlation tensor} by
\bse
\be\label{e:diffusion_tensor}
({D_*}_{\ga\gb})=\begin{pmatrix}
0&0\\
0&2D\, \diff t_*
\end{pmatrix},
\ee
thus,
\be
\lan w_*^\ga(t)\, w_*^\gb(t) \ran={D_*}^{\ga\gb}.
\ee
Additionally defining an {\em \lq inverse\rq\space correlation tensor} by
\be\label{e:inverse_diffusion_tensor}
(\hat{D_*}_{\ga\gb})=\begin{pmatrix}
0&0\\
0&(2D\, \diff t_*)^{-1}
\end{pmatrix},
\ee
\ese
allows us to generalize the distribution of the increments from
Eq. \eqref{e:wiener} as follows
\be\label{e:probability_rest_frame.1}
\mcal{P}_*^{1+1}[w_*^\ga(t)]=
\f{1}{\sqrt{4\pi D\; \diff t_*}}\;
\exp\left[-\f{1}{2}\; \hat{D_*}_{\ga\gb}\,w_*^\ga(t)\,w_*^\gb(t)\right]
\times \gd\left[w_*^0(t)\right].
\ee
Here, the Dirac $\gd$-function on the rhs. accounts for the fact that the
0-component of the stochastic force must vanish in every inertial frame,
comoving  with the Brownian particle at time $t$; compare
Eq. \eqref{e:covariant_stochastic_force}. This also follows more generally
from the identity
\be\label{e:condition.1}
0\equiv\f{\diff}{\diff\tau}(-mc^2)
=m\f{\diff}{\diff\tau}(u_\ga u^\ga)
=2 u_\ga f^\ga,
\ee
which, in the case of the stochastic force, translates to
\be\label{e:condition.2}
0= u_\ga\, w^\ga.
\ee
Hence, we can rewrite the probability distribution \eqref{e:probability_rest_frame.1} as
\be\label{e:probability_rest_frame.2}
\mcal{P}_*^{1+1}[w_*^\ga(t)]=
\f{c}{\sqrt{4\pi D\; \diff t_*}}\;
\exp\left[-\f{1}{2}\; \hat{D_*}_{\ga\gb}\,w_*^\ga(t)\,w_*^\gb(t)\right]
\times
\gd\left[{u_*}_\ga\,w_*^\ga(t)\right],
\ee
where $({u_*}_\ga)=(-c, 0)$ is the covariant 1+1-velocity of the particle
a the comoving rest frame. It should be stressed that, because of the
constraint \eqref{e:condition.2}, only one of the two increments $w^\ga\equiv
\diff W^\ga$ is to be regarded as \lq independent\rq, which is reflected by the appearance of the
$\delta$-function in  \eqref{e:probability_rest_frame.2}. Also note that, due to
the prefactor $c$, the normalization
condition takes the simple form
\be
1=
\left\{\prod_{\ga=0}^1\int_{-\infty}^\infty \diff[w_*^\ga(t)]\right\}
\mcal{P}_*^{1+1}[w_*^\ga(t)].
\ee
Furthermore, analogous to \eqref{e:friction_tensor_covariant}, we have the following more
general representation of the correlation tensors
\bse\label{e:diffusion_tensor_covariant}
\be
D_{\ga\gb}&=& \label{e:diffusion_tensor_covariant_a}
{2D\;\diff\tau}\left(\eta_{\ga\gb}+\f{u_\ga u_\gb}{c^2}\right)\\
\hat D_{\ga\gb} &=& \label{e:diffusion_tensor_covariant_b}
\f{1}{2D\;\diff\tau}\left(\eta_{\ga\gb}+\f{u_\ga u_\gb}{c^2}\right).
\ee
Then, in an arbitrary Lorentz frame, the density \eqref{e:probability_rest_frame.2} can be
written as
\be
\mcal{P}^{1+1}[w^\ga(\tau)]
&=&\notag
\f{c}{\sqrt{4\pi D\; \diff \tau}}\;
\exp\left[-\f{1}{2}\; \hat{D}_{\ga\gb}\,w^\ga(\tau)\,w^\gb(\tau)\right]
\times \gd\left[u_\ga\,w^\ga(\tau)\right]\\
&=&\label{e:RBM_covariant_dens}
\f{c}{\sqrt{4\pi D\; \diff \tau}}\;
\exp\left[-\f{w_\ga(\tau)\,w^\ga(\tau)}{4D\,\diff \tau}\;\,\right]
\times \gd\left[u_\ga\,w^\ga(\tau)\right].
\ee
\ese
To obtain the last line from the first, we have inserted $\hat
D_{\ga\gb}$ from \eqref{e:diffusion_tensor_covariant_b} and then used that $u_\ga
w^\ga=0$, see Eq. \eqref{e:condition.2}.
\par
By virtue of the above results, we are now in the position to write down
the covariant Langevin equations with respect to an arbitrary inertial system:
If a Brownian particle with rest mass $m$, proper time $\tau$ and 1+1-velocity
$u^\gb$ is surrounded by an isotropic, homogeneous heat bath with
constant 1+1-velocity $U^\gb$, then the relativistic Langevin equations of
motions read
\bse\label{e:RBM_covariant}
\be
\diff x^\ga(\tau)&=&\label{e:RBM_covariant.1}
\f{p^\ga(\tau)}{m} \diff \tau\\
\diff p^\ga(\tau)&=&\label{e:RBM_covariant.2} -{\nu^\ga}_\gb
\left[p^\gb(\tau)-m U^\gb\right] \diff \tau +w^\ga(\tau),
\ee
where, according to Eq. \eqref{e:friction_tensor_covariant}, the friction
tensor is given by
\be\label{e:RBM_covariant.3}
{{\nu}^\ga}_\gb=
\nu\left({\eta^\ga}_{\gb}+\f{u^\ga u_\gb}{c^2}\right),
\ee
with $\nu$ denoting the viscous friction coefficient measured in the rest
frame of the particle. This is a first main result of this work. The
stochastic increments $w^\ga(\tau)\equiv \diff W^\ga(\tau)$ are
distributed according to \eqref{e:RBM_covariant_dens} and,
therefore, characterized by
\be
\lan w^\ga(\tau)\ran & =&0,\label{e:RBM_covariant.4} \\
\lan w^\ga(\tau)\, w^\gb(\tau') \ran &=&\label{e:RBM_covariant.5}
\begin{cases}
0, & \tau\ne \tau';\\
{D}^{\ga\gb},& \tau= \tau',
\end{cases}
\ee
\ese
with $D^{\ga\gb}$ given by \eqref{e:diffusion_tensor_covariant_a}.
Note that in each comoving Lorentz frame, in which, at a given moment $t$, the particle is at rest,  the marginal distribution of the spatial momentum increments, defined by
\be\label{e:marginal}
\mcal{P}^{1}[ w(t)]=\int_{-\infty}^\infty \diff[w^0(t)]\;
\mcal{P}^{1+1}[w^\ga(t)],
\ee
reduces to a Gaussian. In the Newtonian limit case, corresponding to $v^2\ll c^2$, one thus recovers
from Eqs. \eqref{e:diffusion_tensor_covariant} and
\eqref{e:RBM_covariant} the usual nonrelativistic Brownian motion.

\subsection{Langevin dynamics in the laboratory frame}
\label{lab}

A laboratory frame $\Sigma_0$ is, by definition, an inertial system, in which
the heat bath is at rest, i.e.,  in $\Sigma_0$ we have $(U^\gb)=(c,0)$ for
all times $t$. Hence, with respect to $\Sigma_0$-coordinates, the two stochastic
differential  Eqs. \eqref{e:RBM_covariant.2} take the form
\bse\label{e:lab-EOM}
\be
\diff {p}&=&\label{e:lab-EOM-a}
-\nu\, {p}\, \diff t+ {w}(t),\\
\diff E &=&\label{e:lab-EOM-b}
-\nu\, {p}{v}\, {\diff t}+ c w^0(t).
\ee
\ese
Here it is important to notice that the stochastic increments $w^\ga(t)$,
appearing on the rhs. of \eqref{e:lab-EOM}, are {\em not} of simple Gaussian
type anymore. Instead, their distribution now also depends on the particle
velocity $v$. This becomes immediately evident, when we rewrite the
increment density \eqref{e:RBM_covariant_dens} in terms of $\Sigma_0$-coordinates. Using
\be
(u_\ga)=\left(-\gc c,\gc v\right),
\qquad\qquad
\gc^{-1}=\sqrt{1-\f{v^2}{c^2}},
\qquad\qquad
(w^\ga)=\left(w^0, w\right),
\ee
we find
\be\label{e:lab_probabilty_density}
\mcal{P}^{1+1}[w^\ga(t)]=
c\left(\f{\gc}{4\pi D\; \diff t}\right)^{1/2}
\exp\left[-\f{w(t)^2-w^0(t)^2}{4D\;\diff t/\gc}
\right]
\times \gd\left[c\gc w^0(t)-\gc v w(t) \right].
\ee
As we already pointed out earlier, the $\delta$-function in
\eqref{e:lab_probabilty_density} reflects the fact that the energy increment $w^0$  is coupled
to the spatial (momentum) increment $w$ {\it via}
\be
0=u_\ga w^\ga=-c\gc w^0+\gc  v  w
\qquad\Rightarrow\qquad
w^0=\f{ v  w}{c}.
\ee
Hence, $w^0$ can be eliminated from the Langevin equations \eqref{e:lab-EOM-b}, yielding
\be
\diff E
=
-\nu\, {p}{v}\, {\diff t}+  v\,  w(t)
=\label{e:lab-EOM-1.2}
 v\,\diff {p}.
\ee
Using the identity
\be\label{e:v-p-identity}
v=
\f{c p}{\sqrt{m^2c^2+{p}^2}},
\ee
we can further rewrite \eqref{e:lab-EOM-1.2} as
\be\label{e:lab-energy}
\diff E=\f{cp}{\sqrt{m^2c^2+{p}^2}}\diff {p}
\qquad \Rightarrow \qquad
E(t)=\sqrt{m^2c^4+{p}(t)^2c^2}.
\ee
Thus, in the laboratory frame $\Sigma_0$ the relativistic Brownian motion is
completely described by the Langevin equation \eqref{e:lab-EOM-a} already. If
we assume that the Brownian particle has fixed initial momentum $p(0)= p_0$ or
initial velocity $v(0)= v_0$, respectively, then the formal solution
of \eqref{e:lab-EOM-a} reads (\cite{VK03} Chap. IX.1)
\be\label{e:labsol-p}
p(t)= p_0 e^{-\nu t}+e^{-\nu t}\int_0^t e^{\nu s} w(s).
\ee
The stochastic process \eqref{e:labsol-p} is determined by the marginal distribution $\mcal{P}^{1}[w(t)]$, defined in Eq.
\eqref{e:marginal}. Performing the integration over the
$\delta$-function in \eqref{e:lab_probabilty_density}, we find
\bse\label{e:lab_density}
\be\label{e:lab_density-1}
\mcal{P}^{1}[w(t)]
&=& \left(\f{1}{4\pi D\gc\; \diff t}\right)^{1/2} \exp\left[
-\f{w(t)^2}{4D\gc\;\diff t} \right],
\ee
where
\be
\gc=
\left[1-\f{v^2}{c^2}\right]^{-1/2}
=\label{e:lab_density-2}
\left[1+\f{p^2}{m^2c^2}\right]^{1/2}.
\ee
\ese
On the basis of Eqs. \eqref{e:lab-EOM-a} and \eqref{e:lab_density} one can
immediately perform computer simulations, provided one still specifies the
rules of stochastic calculus, i.e., which value of $p$ is to be taken to
determine $\gc$ in \eqref{e:lab_density}.  In Sec. \ref{numerics} several numerical results are presented. Before, it is useful to
consider in more detail the Fokker-Planck equations of the
relativistic Brownian motion in the laboratory frame $\Sigma_0$. By
doing so in the next section, it will become clear that, for example, choosing
$p=p(t)$ in Eqs. \eqref{e:lab_density}, would be consistent with an
Ito-interpretation \cite{Ito44,Ito51,VK03} of the stochastic differential
equation \eqref{e:lab-EOM-a}. However, we will also see that alternative
interpretations lead to reasonable results as well.

\section{Derivation of corresponding Fokker-Planck equations}
\label{FPE}

The objective in this part is to derive relativistic Fokker-Planck equations (FPE) for
the momentum density $f(t,p)$ of a free particle in the laboratory frame $\Sigma_0$.
Before we deal with this problem in Sec. \ref{FPE-r}, it is useful to briefly recall the nonrelativistic case.

\subsection{Nonrelativistic case}
\label{FPE-nr}

Consider the nonrelativistic Langevin equation \eqref{e:langevin_b}
\bse\label{e:FPE-nr-LE}
\be
\f{\diff p}{\diff t} = -\nu\, p+ L(t),
\ee
where $p(t)=m v(t)$ denotes the nonrelativistic momentum, and, in
agreement with \eqref{e:langevin_math_density}, the Langevin force
$L(t)$ is distributed according to
\be
\mcal{P}[L(t)]=
\left(\f{\diff t}{4\pi D}\right)^{1/2}
\exp\left[-\f{\diff t}{4D}L(t)^2\right].
\ee
\ese
As is well known \cite{VK03,Schwabl}, the related momentum
probability density $f(t,p)$ is governed by the Fokker-Planck
equation
\be\label{e:FPE-nr}
\f{\p}{\p t} f=
\f{\p}{\p p} \left(\nu p f + D \f{\p}{\p p} f\right),
\ee
whose stationary solution is the Maxwell distribution
\be\label{e:FPE-nr-sol}
f(p)
=\left(\f{\nu}{2\pi D}\right)^{1/2} \exp\left(-\f{\nu p^2}{2D}\right).
\ee

\subsection{Relativistic case}
\label{FPE-r}

We next discuss three different  relativistic Fokker-Planck
equations for the momentum density $f(t,p)$, related to the
stochastic processes defined by \eqref{e:lab-EOM-a} and
\eqref{e:lab_density}.
\par
Our starting point is the relativistic Langevin equation
\eqref{e:lab-EOM-a}, which holds in the laboratory frame $\Sigma_0$
(i.e. in the rest frame of the heat bath). Next we define a new stochastic
process by
\be \label{e:y-definition}
y(t)= \f{w(t)}{\sqrt{\gc}},
\ee
and using \eqref{e:lab_density-2}, we can rewrite \eqref{e:lab-EOM-a} as
\bse\label{e:lab-EOM-y}
\be
\diff {p}&=&\label{e:lab-EOM-y-1} -\nu
\,p\, \diff t+\sqrt{\gc}\; {y}(t),
\ee
where $y(t)$ is distributed according to the momentum-independent density
\be
\mcal{P}_y^1[y(t)] &=&\label{e:lab-EOM-y-2}
\left(\f{1}{4\pi D\; \diff t}\right)^{1/2} \exp\left[ -\f{y(t)^2}{4D\;\diff t}
\right].
\ee
\ese
Thus, instead of the increments $w(t)$, which implicitly
depend on the stochastic process $p$ {\it via} Eqs. \eqref{e:lab_density}, we
 consider ordinary $p$-independent white noise $y(t)$, determined by
 \eqref{e:lab-EOM-y-2}, from now on. Due to the multiplicative
coupling of $y(t)$ in \eqref{e:lab-EOM-y-1}, we must next specify rules  
for the \lq\lq multiplication with white noise\rq\rq [note that, upon
viewing Eqs. \eqref{e:propertime}, \eqref{e:diffusion_tensor_covariant}
and \eqref{e:RBM_covariant} as postulates of the relativistic Brownian motion,
all above considerations remain valid independent of this specification].
\par
In the following subsections, we shall discuss three popular  
multiplication rules, which go back to proposals made by Ito
\cite{Ito44,Ito51,HaTo82,VK03}, by Stratonovich and Fisk
\cite{St64,St66,Fisk63,Fisk65,HaTo82}, and by H\"anggi
\cite{Han78,Han80,Han82} and Klimontovich \cite{Kl94}, respectively. As
well-known \cite{HaTo82,Tr03,VK03}, these different interpretations of the stochastic 
process \eqref{e:lab-EOM-y} result in different Fokker-Planck equations, i.e., the Langevin equation \eqref{e:lab-EOM-y} {\it
  per se} does {\em not} uniquely determine the corresponding  Fokker-Planck
equation; it is the stochastic interpretation of the
multiplicative noise that matters from a physical point of view.
\par
Nevertheless, the three approaches discussed below have in common that,
formally, the related Fokker-Planck equation can be written  as a continuity equation
\cite{Han82}: 
\be\label{e:continuity} \f{\p}{\p t} f(t, p)+\f{\p}{\p
p} j(t, p)=0, 
\ee 
but with different expressions for the probability current $j(t,p)$. It is
worthwhile to anticipate that only for the H\"anggi-Klimontovich approach (see
Sec. \ref{FPE-jutt}) the current $j(t,p)$ takes such a form that the stationary
distribution of \eqref{e:continuity} can be identified with  J\"uttner's relativistic Maxwell distribution \cite{Ju11}.

\subsubsection{Ito approach}
\label{FPE-ito}

According to Ito's interpretation of the Langevin equation
\eqref{e:lab-EOM-y-1}, the coefficient before $y(t)$ is to be evaluated at the
lower boundary of the interval $[t,t+\diff t]$, i.e., we use the pre-point discretization rule
\be\label{e:ito_choice}
\gc=\gc\bigl(p(t)\bigr),
\ee
where as before
$$
\gc(p)=\left(1+\f{p^2}{m^2c^2}\right)^{1/2}.
$$
Ito's choice leads to the following expression for the current
\cite{Ito44,Ito51,HaTo82,VK03}
\be\label{e:ito_current}
j_\mrm{I}(p,t)=-\left[\nu p f+D\f{\p}{\p p} \gc(p) f\right].
\ee
The related relativistic Fokker-Planck equation is obtained by inserting
this current into the conservation law \eqref{e:continuity}. The
current \eqref{e:ito_current} vanishes identically  for
\be\label{e:ito_solution}
f_\mrm{I}(p)=
\f{C_\mrm{I}}{\gc(p)}\exp\left[-\f{\nu}{D}
\int\diff p\, \f{p}{\gc(p)}\right ],
\ee
where $C_\mrm{I}$ is
the normalization constant. Consequently, $f_\mrm{I}(p)$ is a
stationary solution of the Fokker-Planck equation. In view of the
fact that
\be\label{e:integral}
\int\diff p\,
\f{p}{\gc(p)}=c^2m^2\sqrt{1+\f{p^2}{c^2m^2}},
\ee
we find the following explicit representation of \eqref{e:ito_solution}
\be
f_\mrm{I}(p)=
C_\mrm{I}\left(1+\f{p^2}{m^2c^2}\right)^{-1/2} \exp\left(-\gb
\sqrt{1+\f{p^2}{c^2m^2}}\right), \ee where \be \gb=\f{\nu m^2c^2
}{D}.
\ee
The dimensionless parameter $\gb$ can be used to define
the scalar temperature $T$ of the heat bath via the Einstein
relation \be\label{e:temperature} k_B T\equiv \f{mc^2}{\gb}=\f{D}{m
\nu}, \ee with $k_B$ denoting the Boltzmann constant. Put differently, the
parameter $\gb=mc^2/(k_BT)$ measures the ratio between rest mass and
thermal energy of the Brownian particle.

\subsubsection{Stratonovich approach}
\label{FPE-strat}

According to Stratonovich, the coefficient before $y(t)$ in
 \eqref{e:lab-EOM-y-1} is to be evaluated with the mid-point discretization rule, i.e.,
\be
\gc=\gc\biggl(\f{p(t)+p(t+\diff t)}{2}\biggr).
\ee
This choice leads to a different expression for the  current
\cite{St64,St66,Fisk63,HaTo82}, namely
\be\label{e:stratonovich_current}
j_\mrm{S}(p,t)=-\left[\nu p f+D \sqrt{\gc(p)} \f{\p}{\p p}
\sqrt{\gc(p)}\; f\right].
\ee
This Stratonovich-Fisk current $j_\mrm{S}$ vanishes identically for
\be
f_\mrm{S}(p)=\f{C_\mrm{S}}{\sqrt{\gc(p)}}
\exp\left[-\f{\nu}{D} \int\diff p\, \f{p}{\gc(p)}\right ],
\ee
and, by virtue of \eqref{e:integral}, the explicit stationary solution of
 Stratonovich's Fokker-Planck equation reads
\be\label{e:stratonovich_solution}
f_\mrm{S}(p)=
C_\mrm{S} \left(1+\f{p^2}{m^2c^2}\right)^{-1/4}
\exp\left(-\gb \sqrt{1+\f{p^2}{c^2m^2}}\right).
\ee

\subsubsection{H\"anggi-Klimontovich approach}
\label{FPE-jutt}

Now let us still consider the H\"anggi-Klimontovich
stochastic integral interpretation, sometimes  referred to as the transport form
\cite{Han78,Han80,Han82} or also as the kinetic form \cite{Kl94}.
According to this interpretation, the coefficient in front of $y(t)$ in
\eqref{e:lab-EOM-y-1} is to be evaluated at the upper boundary value of the
interval $[t,t+\diff t]$; i.e., within the post-point discretization we set
\be
\gc=\gc\bigl(p(t+\diff t)\bigr).
\ee
This choice leads to the following expression for the current
\cite{Han80,Han82,Kl94}
\be\label{e:K_current}
j_\mrm{HK}(p,t)=-\left[\nu p f+D \gc(p) \f{\p}{\p p} f\right].
\ee
The  current $j_\mrm{HK}$ vanishes identically for
\be
f_\mrm{HK}(p)=
C_\mrm{HK} \exp\left[-\f{\nu}{D} \int\diff p\,\f{p}{\gc(p)}\right ],
\ee
and, by virtue of \eqref{e:integral}, the stationary solution explicitly
reads
\bse \label{e:juettner}
\be\label{e:K_solution}
f_\mrm{HK}(p)=
C_\mrm{HK}\exp\left(-\gb \sqrt{1+\f{p^2}{c^2m^2}}\right).
\ee
Using the temperature definition in \eqref{e:temperature} and the relativistic kinetic energy
formula $E=\sqrt{m^2 c^4+p^2c^2}$, one can further rewrite
\eqref{e:K_solution} in a more concise form as
\be
f_\mrm{HK}(p)= C_\mrm{HK}\exp\left(-\f{E}{k_BT}\right).
\ee
\ese
The distribution function \eqref{e:juettner} is known as the
relativistic Maxwell distribution. It was first obtained by F.
J\"uttner \cite{Ju11} back in 1911. Pursuing a completely different line of
reasoning, he found that \eqref{e:juettner} describes the velocity
distribution of the non-interacting relativistic gas (see also \cite{Sy57}). In contrast to
our approach, which started out  with constructing the relativistic
generalization of the Langevin equations, J\"uttner's derivation
started from a maximum-entropy-principle for the gas.
\par
By comparing \eqref{e:ito_solution}, \eqref{e:stratonovich_solution} and
\eqref{e:K_solution} one readily observes that the stationary solutions
$f_\mrm{I/S}$ differ from the J\"uttner function $f_\mrm{HK}$
through additional $p$-dependent prefactors. In order to illustrate the
differences between the different stationary solutions, it useful to consider
the related velocity probability density functions $\phi_\mrm{I/S/HK}(v)$, which can be
obtained by applying the general transformation law
\be
\phi(v)
\equiv \label{e:relmax}
f(p(v))\left|\f{\p p}{\p v}\right|
\ee
in combination with
$$
p=\f{m v}{\sqrt{1-v^2/c^2}}.
$$
The exponential factor ensures that the velocity density functions $\phi_\mrm{I/S/HK}(v)$ are in fact zero if $v^2>c^2$.
\par
\begin{figure}[h]
\center \epsfig{file=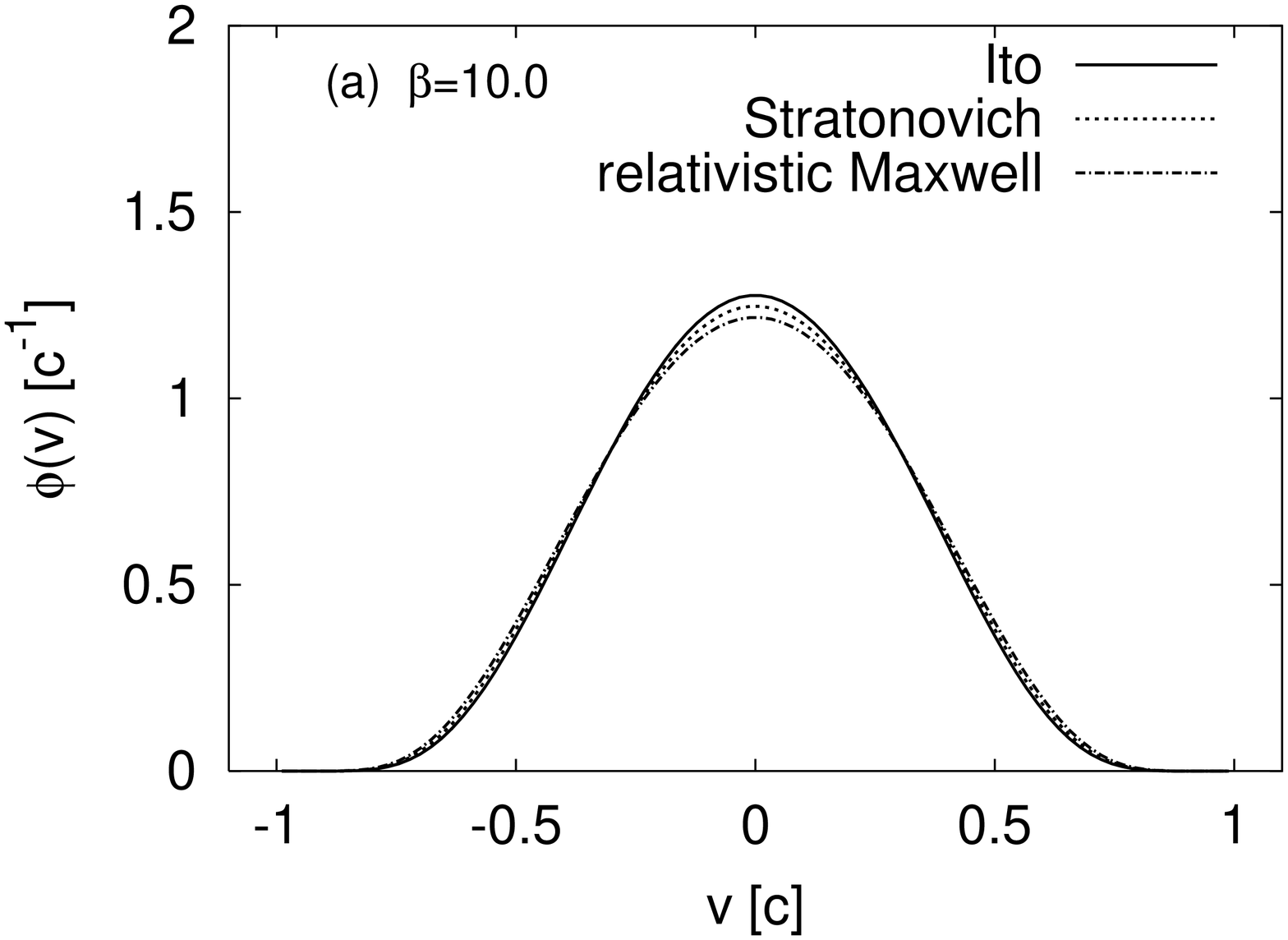,height=5.9cm, angle=-90}
\epsfig{file=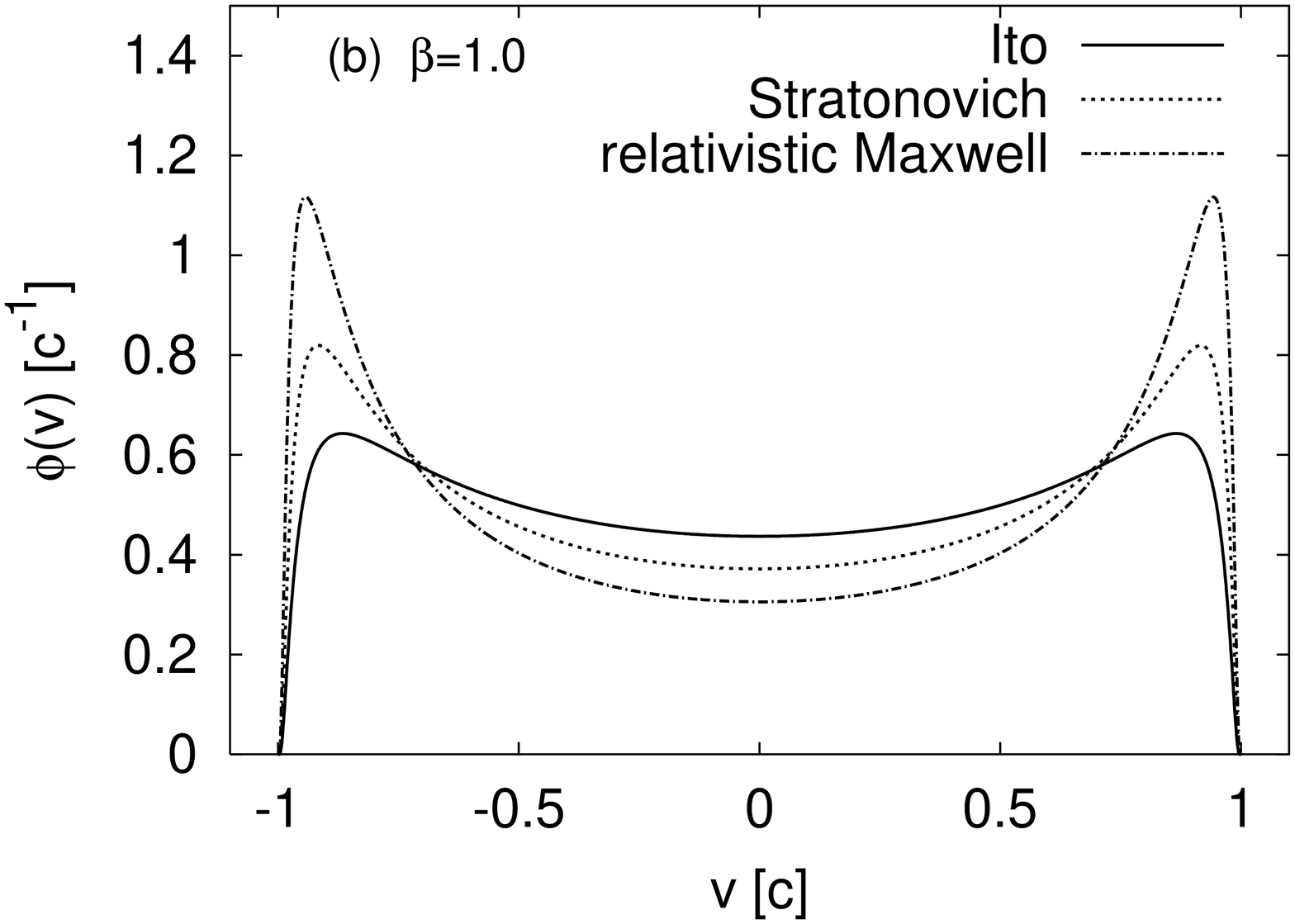,height=5.9cm, angle=-90}
\epsfig{file=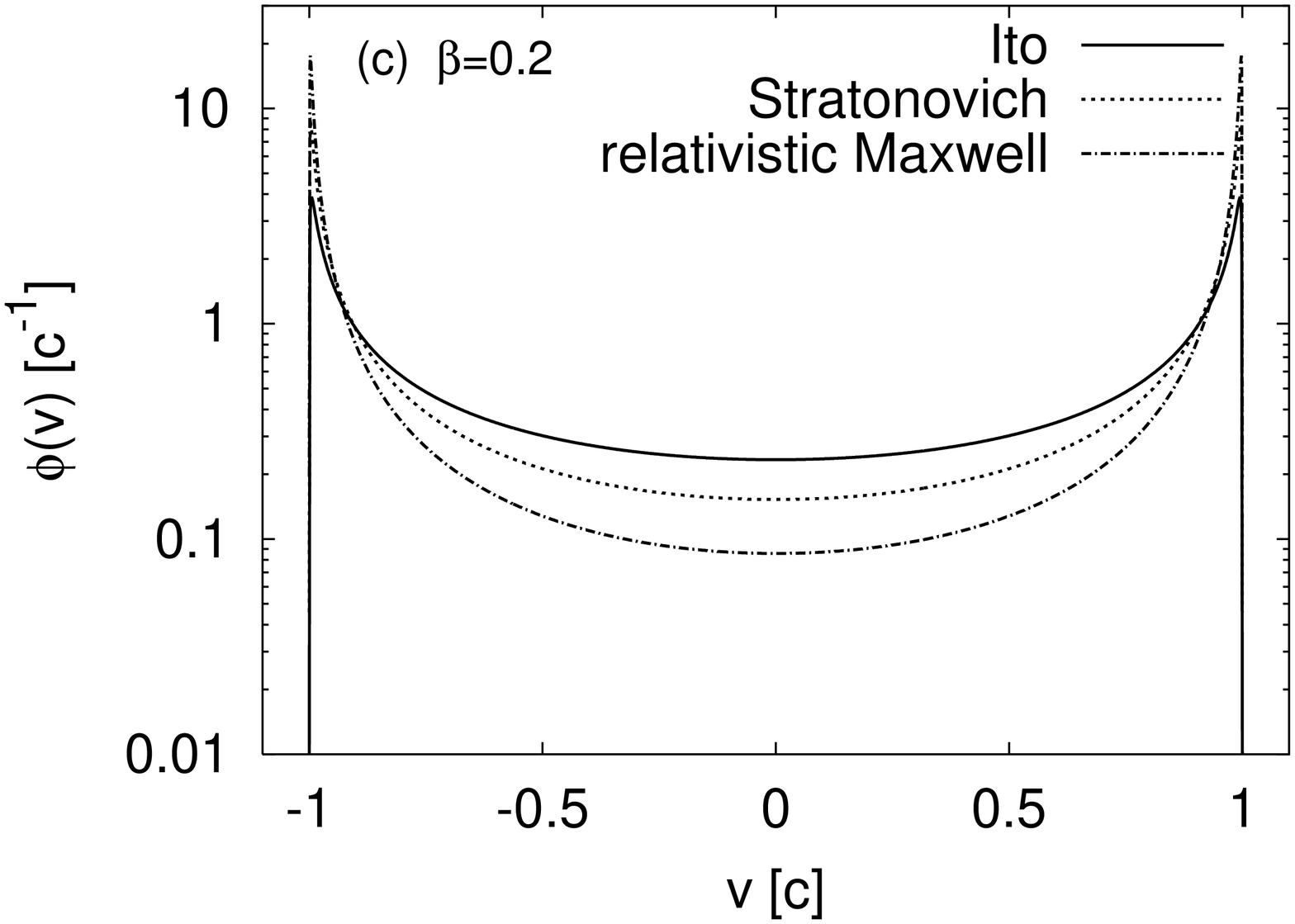,height=5.9cm, angle=-90} \caption{Stationary
solutions $\phi_\mrm{I/S/HK}(v)$ of the relativistic Fokker-Planck
equations, according to Ito (I:solid line), Stratonovich (S:dotted)
and
  H\"anggi-Klimontovich (HK:dashed-dotted). For low temperatures,
  i.e., for $\gb \gg 1$, a Gaussian shape is approached,
  see diagram (a). On the other hand, for very high temperature values,
  corresponding to  $\gb \le 1$, the distributions exhibit a bistable
  shape, and the quantitative deviations between  $\phi_\mrm{I/S/HK}(v)$ increase significantly  as $\gb \to 0$.
\label{fig02}}
\end{figure}

In Fig. \ref{fig02} we have plotted the probability density functions $\phi_\mrm{I/S/HK}(v)$ for
different values of the parameter $\gb$. The normalization constants were
determined by numerically integrating $\phi(v)$ over the interval $[-c,c]$. As
one can observe in diagram  \ref{fig02} (a), for large values of $\gb$,
corresponding to small temperature values $k_B T\ll mc^2$, the  density
functions $\phi_\mrm{I/S/HK}(v)$ approach a common Gaussian shape. On the other hand,
for high temperature values $k_B T\ge mc^2$ the deviations from the Gaussian
shape become essential. The reason is that, for a (virtual) Brownian ensemble in the
high-temperature regime, the majority of particles assumes velocities, that are
close to the speed of light. It is also clear that in other Lorentz frames
$\Sigma'$, which are not rest frames of the heat bath, the stationary
distributions will no longer stay symmetric around $v=0$. Instead, they will
be centered around the non-vanishing $\Sigma'$-velocity $V'$ of the heat bath.
\par
As an obvious question then arises, which of the above approaches (Ito,
Stratonovich or H\"anggi-Klimontovich)  is the physically correct
one. We believe that, at this level of analysis, it is impossible to
provide a definite answer to this question. Most likely,  the answer
to this problem requires additional information about the
microscopic structure of the heat bath (see, e.g., the discussion of
Ito-Stratonovich dilemma in the context of \lq
internal/external\rq\space noise as given in Chap. IX.5 of van
Kampen's textbook \cite{VK03}). At this point, it might be
worthwhile to mention that the relativistic Maxwell distribution
\eqref{e:juettner} is also obtained via the transfer probability
method used by Schay, see Eq. (3.63) and (3.64) in
Ref. \cite{Sc61}, and that this distribution also results in the
relativistic kinetic theory \cite{DG80}.
By contrast, the recent work of Franchi and Le Jan \cite{FrLJ04} is
based on the Stratonovich approach. From pure physical insight,
however, it is the transport form interpretation of H\"anggi and 
Klimontovich that is expected to provide the physical correct
description.

\section{Numerical investigations}
\label{numerics}

The numerical results presented in this section were obtained on the basis of
the relativistic Langevin equation \eqref{e:lab-EOM-y}, which holds in the
laboratory frame $\Sigma_0$. For simplicity, we confined ourselves here to
considering the  Ito-discretization scheme with fixed time step $\diff t$
, see Sec. \ref{FPE-ito}. In all simulation we have used an ensemble size
of $N=10000$ particles. Moreover, a characteristic unit system was fixed by
setting $m=c=\nu=1$. Formally, this corresponds to using re-scaled dimensionless
quantities, such as $\tilde{p}=p/mc$, $\tilde{x}=x \nu/c$, $\tilde{t}=t\nu$,
$\tilde{v}=v/c$  etc.. The simulation time-step was always chosen as $\diff t
=0.001 \nu^{-1}$, and the Gaussian random variables $y(t)$ were generated by
using a standard random number generator.

\subsection{Distribution functions}
\label{distributions}

\begin{figure}[h]
\center
\epsfig{file=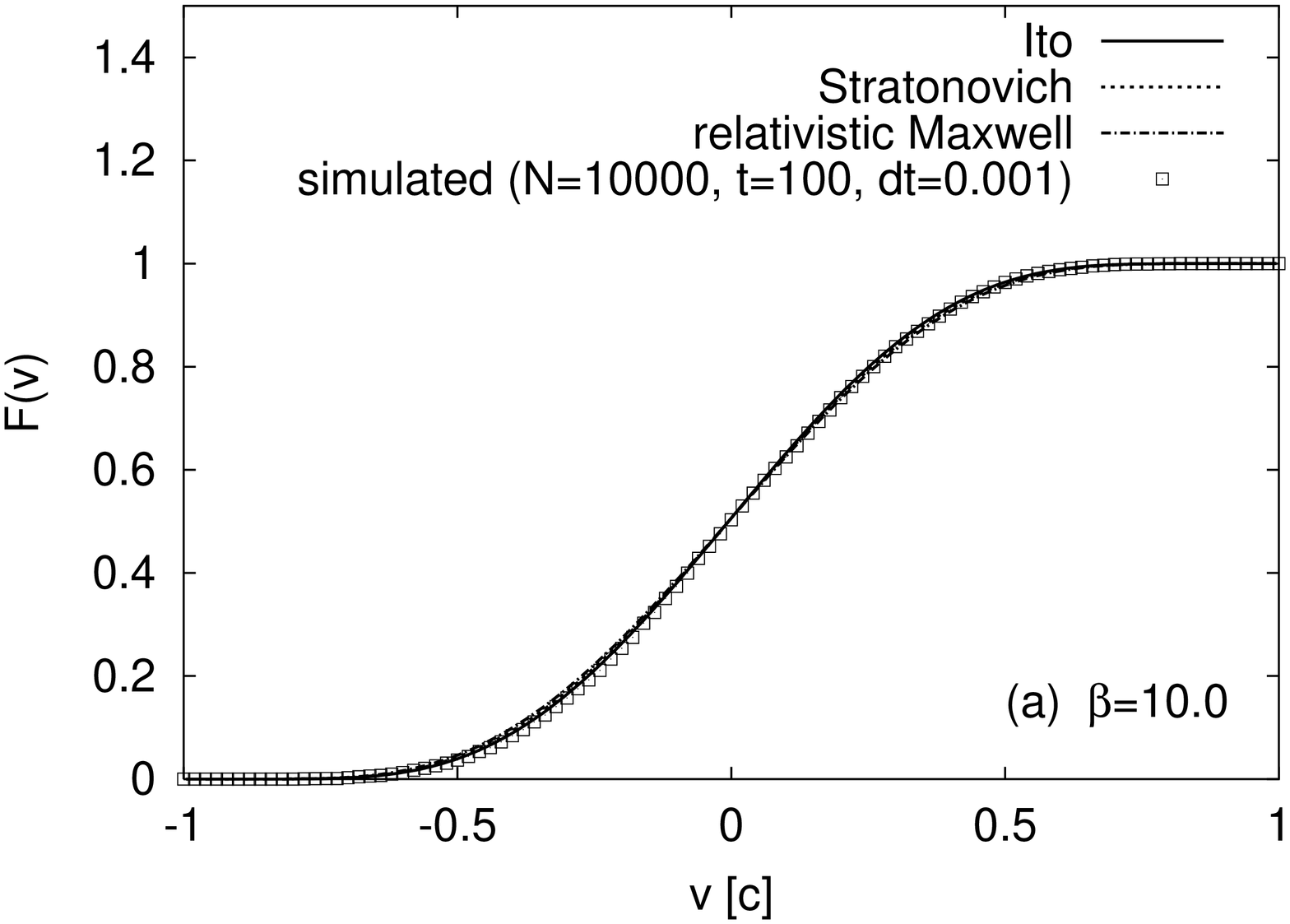,height=5.9cm, angle=-90}
\epsfig{file=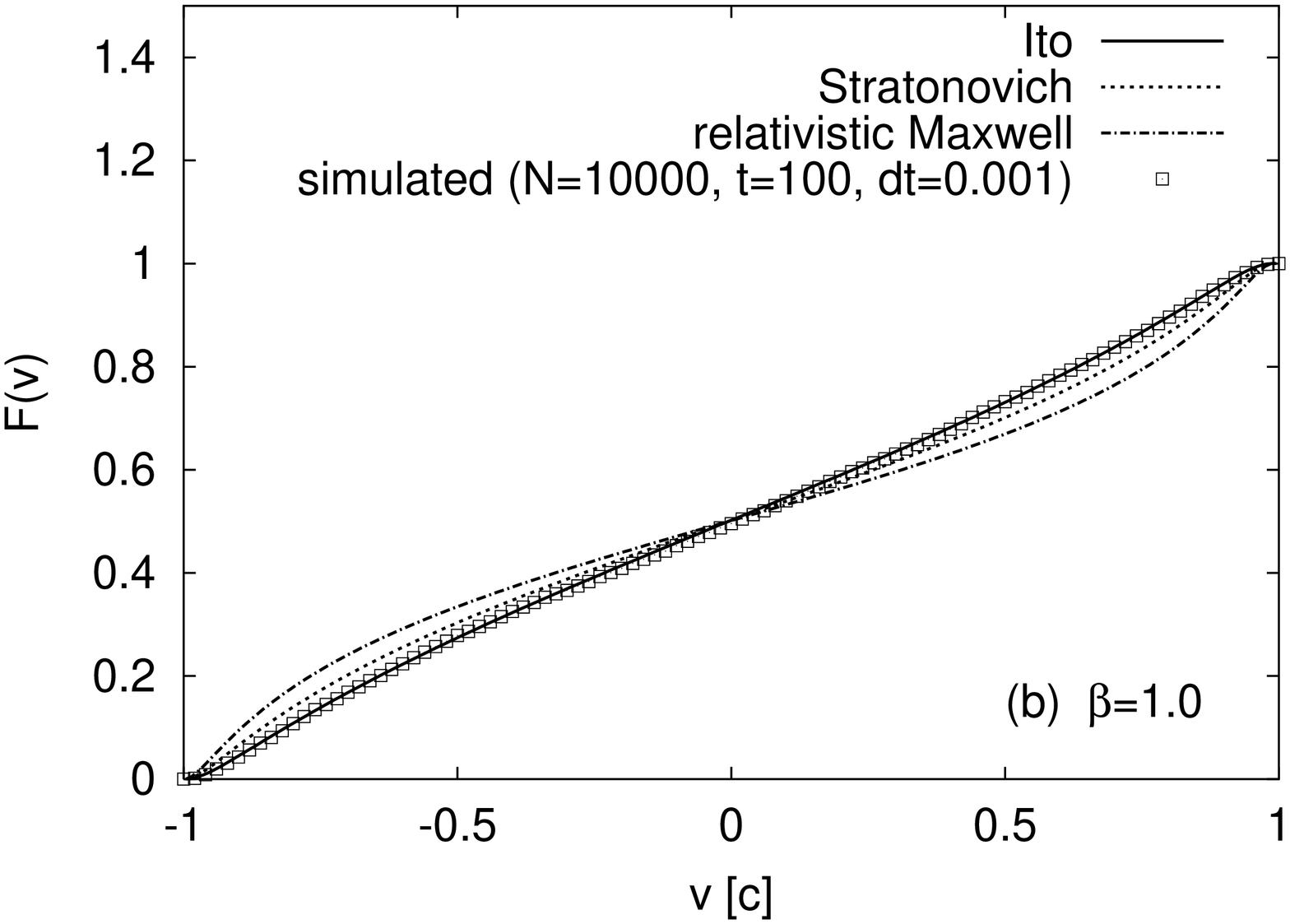,height=5.9cm, angle=-90}
\epsfig{file=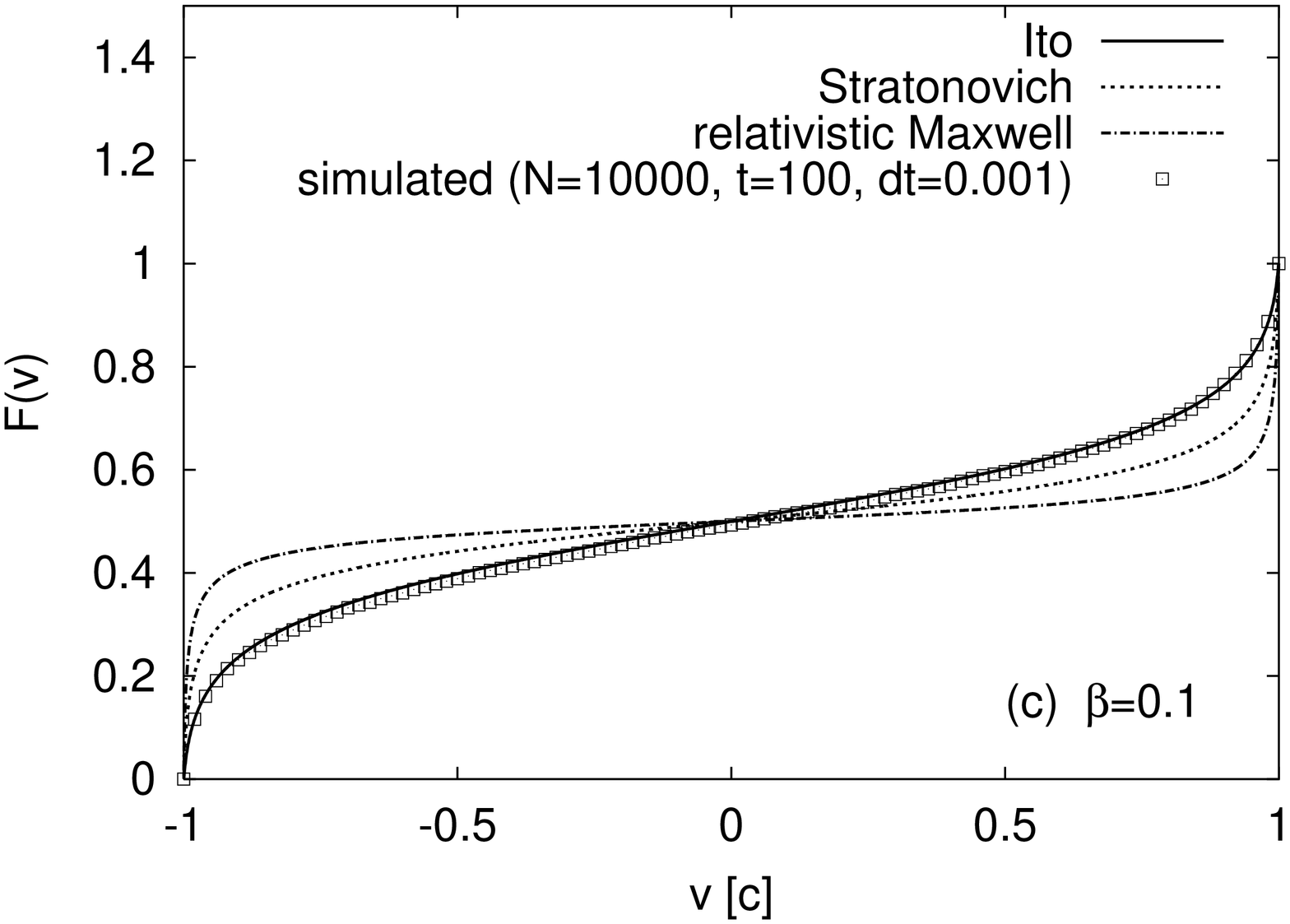,height=5.9cm, angle=-90}
\caption{These diagrams show a comparison between numerical and analytical
  results for the stationary cumulative distribution function $F(v)$ in the laboratory frame
  $\Sigma_0$. (a) In the nonrelativistic limit $\gb\gg 1$ the stationary
  solutions of the three different FPE are nearly indistinguishable. (b-c) In
  the relativistic limit case $\gb \le 1$, however, the stationary solutions exhibit
  deviations from each other. Because our simulations are based on an
  Ito-discretization  scheme, the numerical values (squares) are best
  fitted by the Ito solution (solid line). \label{fig03}}
\end{figure}

In our simulations we have numerically measured the cumulative velocity distribution
function $F(t,v)$ in the laboratory frame $\Sigma_0$. Given the probability density
$\phi(t,v)$, the cumulative velocity distribution function is defined by
\be\label{e:distribution_function}
F(t,v)=\int_{-c}^v\diff u\; \phi(t,u).
\ee
In order to obtain $F(t,v)$ from numerical simulations, one simply measures
the relative fraction of particles with velocities in the interval
$[-c,v)$. Figure \ref{fig03} shows the numerically
determined {\it stationary}\/ distribution functions (squares), taken at
time $t=100 \nu^{-1}$  and also the corresponding analytical curves
$F_\mrm{I/S/HK}(v)$. The latter were obtained by numerically integrating the
formula \eqref{e:distribution_function}, using the three different stationary
density functions $\phi_\mrm{I/S/HK}(v)$ from Sec. \ref{FPE}.
\par
As one can see in diagram \ref{fig03} (a), for low temperature values,
corresponding to  $\gb\gg 1$, the three stationary distribution functions are
nearly indistinguishable. For high temperatures, corresponding to $\gb
\le 1$, the stationary solutions exhibit significant quantitative
differences, see Fig. \ref{fig03} (b) and (c). Since our simulations are based
on an Ito-discretization  scheme, the numerical values (squares) are best
  fitted by the Ito solution (solid line). Also note that the quality of the fit
  is very good for the parameters chosen in the simulations, and that this
  property is conserved over several magnitudes of $\beta$. This suggests that
  numerical simulations of the Langevin equations provide a very useful tool, if
  one wishes to study relativistic Brownian motions in more complicated settings, e.g.,
  in higher dimensions or in the presence of additional external fields
  and interactions. In this context, it should again be stressed that the
  appropriate choice of the discretization rule is especially important in
  applications to realistic systems.

\subsection{Mean square displacement}
\label{mean_square_displacement}

In this part we consider the spatial mean square displacement of the free {\it
  relativistic} Brownian motion. Since this quantity is easily accessible in
experiments, it has played an important role in the verification of the
nonrelativistic theory.
\par
As before, we consider an ensemble of $N$ independent Brownian particles with coordinates
$x_{(i)}(t)$ in $\Sigma_0$ and initial conditions $x_{(i)}(0)=0,v_{(i)}(0)=0$
for $i=1,2,\ldots,N$. The position mean value is defined as
\be
\ovl{x}(t)\equiv \f{1}{N}\sum_{i=1}^N x_{(i)}(t),
\ee
and the related second moment is given by
\be
\ovl{x^2}(t)\equiv \f{1}{N}\sum_{i=1}^N \left[x_{(i)}(t)\right]^2.
\ee
The empirical mean square displacement can then be defined as follows
\be
\sigma^2(t)\equiv \ovl{x^2}(t)-\left[ \ovl{x}(t)\right]^2.
\ee
Cornerstone results in the nonrelativistic theory of the one-dimensional
Brownian motion are
\bse
\be
\lim_{t\to +\infty} \ovl{x}(t)&\to&\label{e:mean-a} 0,\\
\lim_{t\to +\infty} \f{\sigma^2(t)}{t}&\to& 2D^x,
\ee
\ese
where the constant
\be\label{e:diffusion}
D^x=\f{k_B T}{m\nu}=\f{D}{m^2\nu^2}
\ee
is the nonrelativistic coefficient of diffusion in coordinate space (not to be
confused with noise parameter $D$).
\par
\begin{figure}[h]
\center
\epsfig{file=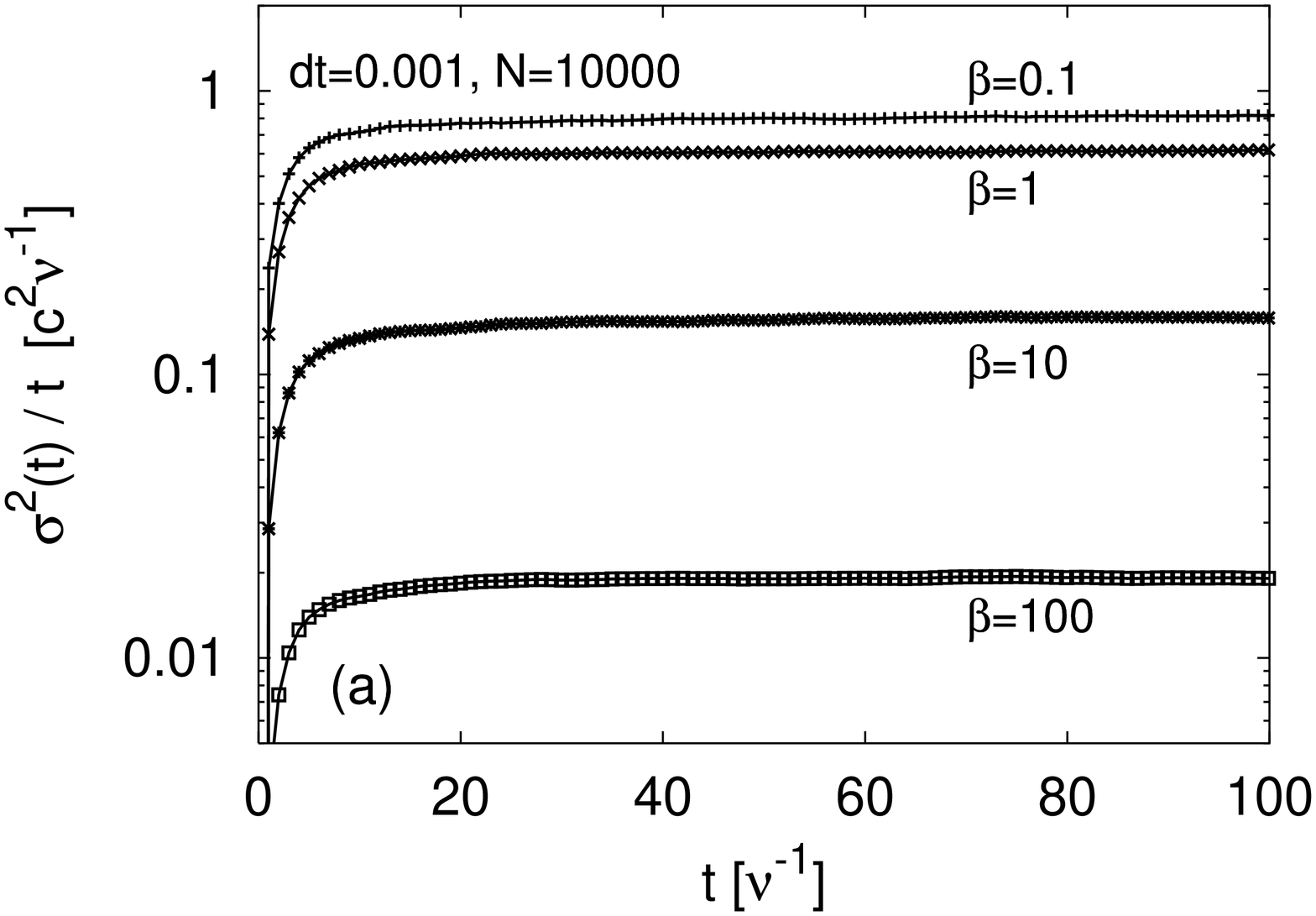,height=6cm, angle=-90}
\epsfig{file=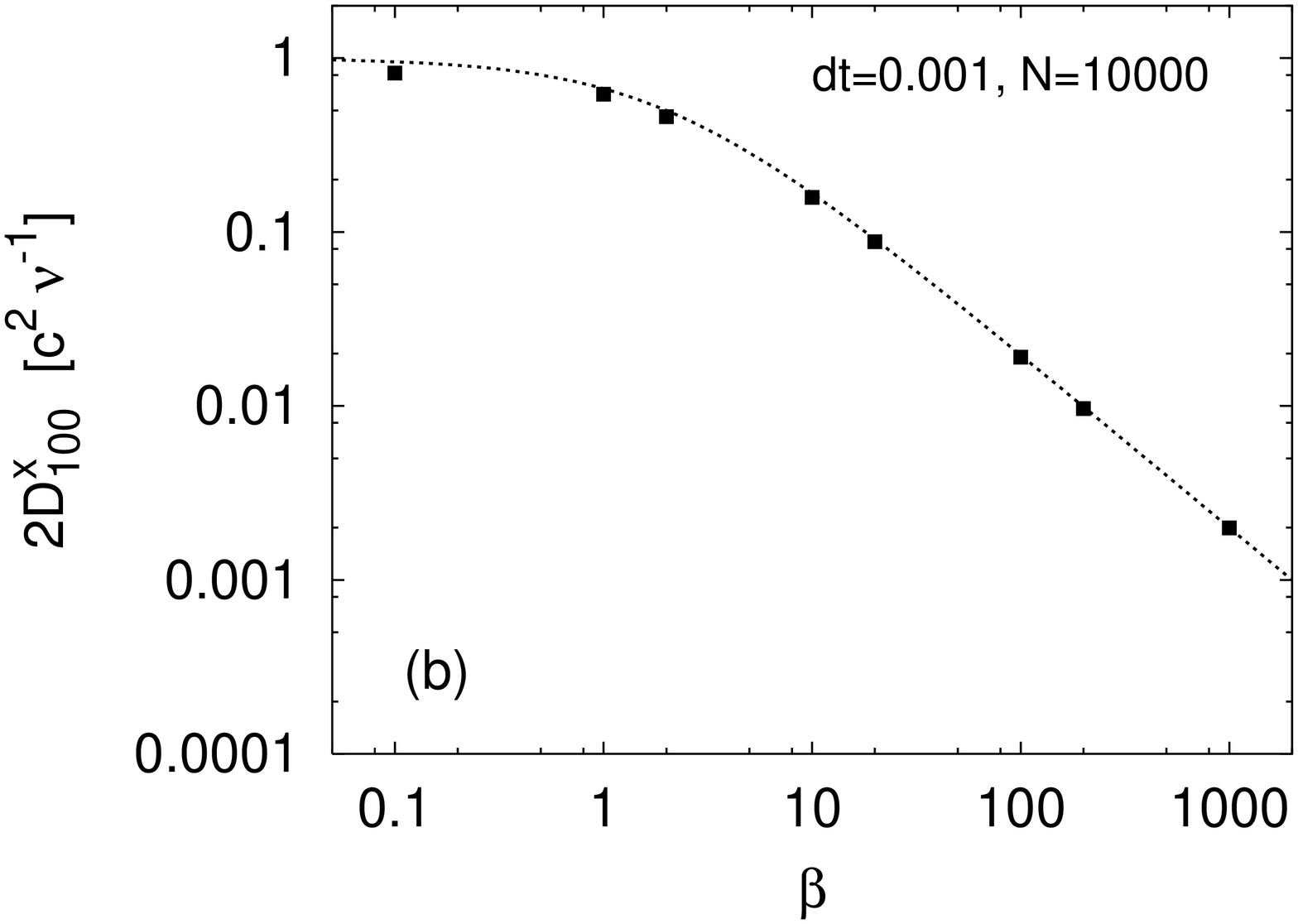,height=6cm, angle=-90}
\caption{(a) Mean square displacement, divided by time $t$, as numerically
  calculated for different $\gb$-values in the laboratory frame
  $\Sigma_0$ (rest frame of the heat bath). As evident from this diagram, for
  the relativistic Brownian motion the related asymptotic mean square
  displacement grows linearly with $t$. (b) The coordinate space diffusion
  constant  $D^x_{100}(\gb)$ was numerically determined at time $t=100
  \nu^{-1}$. The dashed line corresponds to the empirical fitting formula
  $D^x(\gb)=c^2\nu^{-1}(\gb+2)^{-1}$, which reduces to the
  classical nonrelativistic result  $D^x\simeq  c^2/(\nu\gb) =kT/(m\nu)$ for $\gb \gg 2$.
\label{fig04}}
\end{figure}

It is therefore interesting to consider
the asymptotic behavior of the quantity $\sigma^2(t)/t$ for relativistic
Brownian motions, using again the Ito-relativistic Langevin dynamics from Sec. \ref{FPE-ito}.
In Fig. \ref{fig04} (a) one can see the corresponding
numerical results for different values of $\gb$. As one can observe in this
diagram, for each value of $\gb$, the quantity $\sigma^2(t)/t$ converges to a
constant value. This means that, at least in the laboratory frame $\Sigma_0$, the
asymptotic mean square displacement of the free relativistic Brownian motions increases linearly with
$t$. For completeness, we mention that according to our simulations the
asymptotic relation \eqref{e:mean-a} holds in the relativistic case, too.
\par
In spite of these similarities between nonrelativistic and relativistic
theory, an essential difference consists in the explicit temperature
dependence of the limit value $2D^x$. As illustrated in Fig.  \ref{fig04} (b),
the numerical limit values $2D^x_{100}$, measured at time $t=100 \nu^{-1}$, are
well fitted by the empirical formula
\be\label{e:fitting}
D^x=\f{c^2}{\nu (\gb+2)},
\ee
which reduces to the nonrelativistic result \eqref{e:diffusion} in the limit
case $\beta\gg 2$ (low-temperature limit case).
We will leave it as an open problem here, to find an analytical justification
  for the empirically determined formula \eqref{e:fitting}. Instead we merely mention that, upon noting \eqref{e:v-p-identity}, the
relativistic Fokker-Planck equations for the full phase space
density reads
\be\label{e:FPE-rel-ex1}
\f{\p}{\p t}f(t, p, x)+\f{cp}{\sqrt{m^2c^2+ p^2}} \f{\p}{\p x} f(t, p, x)
=
- \f{\p}{\p p}j_\mrm{I/S/HK}(t,p,x) ,
\ee
which might serve as a suitable starting point for such an analysis. Compared with the relativistic
Fokker-Planck equations from the preceding section, the second term
on the lhs. of \eqref{e:FPE-rel-ex1} is new. In particular, we
recover the relativistic Fokker-Planck equations for the marginal density
  $f(t,p)$, see Sec. \ref{FPE}, by integrating Eq. \eqref{e:FPE-rel-ex1} over a
spatial volume with appropriate boundary conditions. Finally, we
mention once again that also \eqref{e:FPE-rel-ex1}, as well as all
the other results that have been presented in this part, exclusively
refer to the laboratory frame $\Sigma_0$.

\section{Conclusion}
\label{summary}

Concentrating on the simplest case of 1+1-dimensions, we have put forward the Langevin dynamics
for the stochastic motion of free relativistic Brownian particles in a
viscous medium (heat bath). Analogous to the nonrelativistic
Ornstein-Uhlenbeck theory of Brownian motion \cite{UhOr30,HaJu95,HaTo82,VK03}, it was
assumed that the heat bath can, in good approximation, be regarded as
homogenous. Based on this assumption, a covariant generalization
of the Langevin equations has been constructed in
Sec. \ref{langevin_approach}. According to these generalized stochastic
differential equations, the viscous friction between Brownian particle and
heat bath  is modeled by a friction tensor $\nu_{\ga\gb}$. For a homogeneous
heat bath this friction tensor has the same structure as the pressure tensor
of a perfect fluid \cite{Weinberg}. In particular, it is uniquely determined
by the value of the (scalar) viscous friction coefficient $\nu$, measured in
the instantaneous rest frame of the particle
(Sec. \ref{viscous_friction}). Similarly, the amplitude of the stochastic
force is also governed by a single parameter $D$, specifying the Gaussian
fluctuations of the heat bath, as seen in the instantaneous rest frame of the
particle (Sec. \ref{stochastic_force}).

\par
In Sec. \ref{lab} the relativistic Langevin equations have been derived in
special laboratory coordinates, corresponding to a specific class of Lorentz
frames, in which the heat bath is assumed to be at rest (at all times). One finds that the corresponding relativistic distribution
of the momentum increments now also depends on the history of the momentum
coordinate. This fact is in contrast with the properties of ordinary Wiener
processes \cite{Wi23,KaSh91}, underlying nonrelativistic standard Brownian
motions with \lq additive\rq\space Gaussian white noise. However, as shown in Sec. \ref{FPE}, it is possible to find an
equivalent Langevin equation, containing \lq multiplicative\rq\space Gaussian white noise.
\par

In order to achieve a more complete picture of the relativistic Brownian
motion, the corresponding relativistic Fokker-Planck equations (FPE) have been discussed in
Sec. \ref{FPE} (again with respect to the laboratory coordinates with the heat bath at rest).
Analogous to nonrelativistic processes with \lq
multiplicative\rq\space noise, one can opt for different interpretations of the
stochastic differential equation, which result in different FPE. In this
paper, we concentrated on the three most popular cases, namely the Ito, the
Stratonovich-Fisk, and the H\"anggi-Klimontovich interpretation. We discussed
and compared the corresponding stationary solutions for a free Brownian
particle. It could be established that only the H\"anggi-Klimontovich
interpretation is consistent with the relativistic Maxwell distribution. This
very distribution was first derived by J\"uttner \cite{Ju11} as the
equilibrium velocity distribution of the relativistic ideal gas. Later on, it
was also discussed by Schay in the context of relativistic diffusions
\cite{Sc61} and by de Groot {\it et al.} in the framework of the relativistic
kinetic theory \cite{DG80}.
\par
In Sec. \ref{numerics} we presented numerical results, obtained on the basis
of an Ito pre-point discretization rule. The simulations indicate that
-- analogous to the nonrelativistic case -- the relativistic mean square
displacement grows linearly with the laboratory coordinate time; the
temperature dependence of the related spatial diffusion constant, however,
becomes more intricate. In principle, the numerical results suggest that
simulations of the Langevin equations may provide a very useful tool for
studying the dynamics of relativistic Brownian particles. In this context it
has to be stressed that an appropriate choice of the discretization rule is
especially important in applications to realistic physical systems. If, for
example, agreement with the kinetic theory \cite{DG80} is desirable, then a post-point
discretization rule should be used.
\par
From the methodical point of view, the systematic relativistic Langevin
approach of the present paper differs from Schay's transition probability
approach \cite{Sc61} and also from the techniques applied by other authors
\cite{OrHo03,Du65,BY81}. As we shall discuss in a forthcoming
contribution, the above approach can easily be generalized to settings which
are more relevant with regard to experiments (such as the 1+3-dimensional
case, the presence of additional external force fields, etc.).
\par
With regard to future work, several challenges remain to be solved. For
example, one should try to derive an analytic expression for the
temperature dependence of the spatial diffusion constant. A suitable
starting point for such studies might be the FPE for the full phase space
density given in Eq. \eqref{e:FPE-rel-ex1}. Another possible task consists in finding
explicit exact or at least approximate time-dependent solutions of
the relativistic FPE. Furthermore, it seems also interesting to
consider extensions to general relativity, as, to some extent,
recently discussed in the mathematical literature \cite{FrLJ04}.  In
this context,  the physical consequences of the different
interpretations (Ito {\it vs.} Stratonovich {\it vs.}
H\"anggi-Klimontovich) become particularly interesting.

\begin{acknowledgments}
J. D. would like to thank S. Hilbert, L. Schimansky-Geier and
S. A. Trigger for helpful discussions.
\end{acknowledgments}


\bibliography{RBM,RelKin,StochCalc,FokkerPlanck}

\end{document}